\documentclass[
  oneside,
  11pt, a4paper,
  footinclude=true,
  headinclude=true,
  cleardoublepage=empty
]{article}
\usepackage[left=2cm,right=2cm,top=2cm,bottom=2cm]{geometry}

\usepackage{amsmath}
\usepackage{amssymb}
\usepackage{amsthm}
\usepackage{acronym}
\usepackage{braket}
\usepackage{array}
\usepackage{bm}

\usepackage[nottoc, notlof, notlot]{tocbibind}

\usepackage[T1]{fontenc}
\usepackage[utf8]{inputenc}

\usepackage{graphicx}

\usepackage{cancel}
\usepackage{ulem}

\usepackage{pgf, tikz}

\usepackage{aeguill} 
\usepackage{hyperref} 
\usepackage{pdfpages} 

\usepackage{comment} 

\usepackage{mathtools}

\newtheorem{lemma}{Lemma}
\newtheorem{note}{Note}

\title{Delta-Bose Gas: the Matrix Elements of the Particle Number Operator as a Determinant}

\usepackage{authblk}

\author[1]{Benoit Vallet} 
\affil[1]{Institut de Physique Théorique, Université Paris Saclay, CEA, CNRS, F-91191 Gif-sur-Yvette, France}

\usepackage{mdframed}
\newmdtheoremenv{fthm}{Theorem}

\begin{document}

\maketitle
\vspace{3cm}
\begin{abstract}
We first propose a pedestrian review of the Coordinate Bethe Ansatz for the $\delta$-Bose gas on an infinite axis. The scalar product of Bethe states, briefly reviewed, provides the first keystone to compute a compact expression for the Matrix Elements of the Particle Number Operator as conjectured by V. Terras, the main result of the following reflection.
\end{abstract}
\vspace{3cm}

\newpage
\tableofcontents

\newpage

\section{Introduction}
The $\delta$-Bose gas on an infinite axis provides an excessively simple integrable model: the coordinate Bethe ansatz for the $\delta$-Bose gas incarnate the most primitive realization of the Bethe-Ansatz machinery, while the absence of boundary wall or cyclic conditions in this system makes the Bethe equations to vanish. 

It thus constitutes a logical first step to address interesting aspects featuring some class of integrable systems before eventually studying these in a more demanding frame. 

We are here first interested, from pedagogical motivations, in understanding how does some determinants, namely the Izergin-Korepin and Slavnov determinants, appear and behave in various domains of the integrability world. 
\\

We shall in those notes depict a reasoning and computations, in the frame of the infinite $\delta$-Bose gas, that will lead us to a compact expression for the Matrix Elements of the Particle Number Operator (MEPNO) in term of a determinant, as conjectured by V. Terras \cite{unpublishedTP}.  

While the ansatz is friendly, the arising computations have to be ran with caution. To this end the rest of the introduction is devoted to the fixing of some conventions, the definition of the analytic objects of recurrent use, and exposition of the main result of this reflection, that is a compact expression for the MEPNO.
\\

Follows the core part of these notes, organized as follows:
\\
In Section \ref{SectDBGandSP} the spectral problem of the $\delta$-Bose gas is presented, and solved by means of the Coordinate Bethe Ansatz. 
\\
An expression for the scalar product in terms of the Izergin-Korepin determinant is reached using a useful lemma due to Gaudin, this latter playing an important role in the reflections to come after. 
\\
This whole section follows the appaoch of the problem as described by Gaudin \cite{Gaudin}. 

Then the MEPNO is computed in Section \ref{SectMEPNO}, and its expression in terms of a determinant rigorously obtained. 

While the key steps are reviewed in this latter section, their detailed understanding will require to visit the Appendix, in which pen and paper are of true support.

\subsection{Conventions and notations}

The calculations to come will involve numerous manipulations over sets of variables, as well as products of functions over their elements. It will then be very convenient to introduce some notations to keep our expressions as sober and intelligible as possible. 
\\

$\bullet$ For $\bar\alpha$ a set of parameters, its $j^{th}$ element will be written $(\bar\alpha )_j$, or simply $\alpha_j$ when no confusion is possible, and its cardinal $\#\bar\alpha$. 
\\

$\bullet$ The ordering of such a set will play an essential role in our development, as sums over these ordering are to appear. 

The different ordering will be generated by action of permutations. A permutation will act on a set by reordering its elements: for $P\in S_{n}$ a permutation and a set $\bar\alpha=\{ \alpha_1,\cdots ,\alpha_n \}$, we define the set $\bar P\bar\alpha$ by its elements $\bar P\bar\alpha=\{ \alpha_{P^{-1}1},\cdots , \alpha_{P^{-1}n} \}$. In other words we define the $i^{th}$ element of $\bar\alpha$ to be the $Pi^{th}$ element of $\bar P\bar\alpha$: a permutation act on the ordering of the elements of a set, not fundamentally on their labels.  

This definition, which may seem unnatural, ensures the associativity of this action:  $\overline{Q}( \overline{P}\overline{\alpha})=\overline{QP}\overline\alpha$. 
\\

\paragraph{Example.}
Consider a set of three elements $\bar\alpha=\{a,b,c \}$ and the cyclic permutation $P=(123)=(12)(23)$ ($(ij)j=i$).

Thus we have the permuted sets $\overline{P} \bar\alpha=\{ c,a,b \}$, $\overline{P^2} \bar\alpha=\{ b,c,a \}$ and $\overline{P^3} \bar\alpha=\{ a,b,c \}=\bar\alpha$.
\vspace{1cm}

$\bullet$ It is now natural to define the class of equivalence $\mathcal{C}(\bar\alpha)$ of a set $\bar\alpha$ by the ensemble of sets related by permutations containing $\bar\alpha$: $\mathcal{C}(\bar\alpha)=\{ \bar P\bar\alpha, P\in S_n \}$. A class of equivalence only contains information about the content of its attached sets, without regards to their ordering. 

As sums over classes of equivalence of partitions will appear in the following, it will be convenient to define the representative element of a class $\mathcal{C}(\bar\alpha)$, referred to as the \textit{normal ordering} of $\bar\alpha$, denoted $:\bar\alpha:$ . By normal ordering we thus consider the choice, arbitrary and unspecified, of a particular element of any class of equivalence.  
\\

$\bullet$ By feeding a function\footnote{for functions initially defined for single arguments, i.e. not fundamentally depending on sets of variables.} with a set we express the product of functions:
$f(\bar\alpha,\bar\beta)=\prod_{i,j}f(\alpha_i,\beta_j)$. 

It will as well sometimes be necessary to operate these product over ordered indexes:
For $\sim$ a relation of order,
$f^\sim (\bar\alpha,\bar\beta)\equiv\prod_{i\sim j}f(\alpha_i,\beta_j)$.
\\

$\bullet$ We define the scalar product of two sets $\bar{u}$ and $\bar{x}$ of lenghts $M$ by $(\bar{u},\bar{x}) = \sum_{i=1}^M u_{i}x_i$. Note that for any permutation $P\in S_M$ we have $(\bar{u},\bar{x})=(\bar P\bar{u},\bar P\bar{x})$. 
\\

$\bullet$ For $u$ and $v$ two variables, we define the rational functions
\begin{align}
\label{RatFct1}
f(u,v) & =1+g(u,v)\equiv 1+\frac{ic}{u-v}
\\
h(u,v) & =\frac{f(u,v)}{g(u,v)}  =\frac{u-v+ic}{ic} 
\\
\label{RatFct2}
t(u,v) & = \frac{g(u,v)}{h(u,v)}=\frac{(ic)^2}{(u-v)(u-v+ic)}=\frac{ic}{u-v}-\frac{ic}{u-v-ic}
\end{align}
which are the analytic building blocs of the objects of concern in the following. These are not uniquely linked to the problem we are to consider, namely the $\delta$-Bose gas, but more generally to one could call \textit{rational} integrable systems.
 \\
 
$\bullet$ We also define, for a set $\bar{u}$, the function
\begin{align}
F(\bar{u})=f^<(\bar{u},\bar{u})=  \prod_{i<j}\left(1+\frac{ic}{u_{i}-u_{j}}  \right)  ,\quad c\in \mathbb{R}
\end{align} 
that will be introduced while diagonalizing our Hamiltonian.
\\

\begin{note}
For $c\in\mathbb{R}$ we have $F^\star(\bar{u})=F(\bar T\bar{u})$, where we define the mirror permutation 

$Ti=\#\bar{u}-i+1,\ i\in\{1,\cdots , \#\bar{u}\}$.
\end{note}

\subsection{Definitions}
We now have to define several objects that will be of recurrent use in the following. 
\\

$\bullet$ For two sets of parameter $\bar{u}$ and $\bar{v}$ of equal length, $\#\bar{u}=\#\bar{v}=n$, we define the Izergin-Korepin determinant

\begin{align}
\nonumber
\mathcal{K}_n(\bar{u}|\bar{v}) \equiv & g^<(\bar{u},\bar{u})\ g^>(\bar{v},\bar{v})\ h(\bar{u},\bar{v})\ \textrm{det}_{i,j}[t(u_i,v_j)]
\\ \label{IKDef}
= & \textrm{det}_{i,j}[h^{-1}(u_i,v_j)] \textrm{det}_{i,j}[t(u_i,v_j)]
\end{align}
where we used the well know factorized expression for the Cauchy determinant in the last line. 

Note that $g(u,v)=-g(v,u)$. This anti-symmetry, alongside the anti-symmetry of the determinant under transposition of lines or columns, makes this object totally symmetric under permutation inside each family of parameters: 
\begin{align}
\label{KSym}
\mathcal{K}_n(\bar{u}|\bar{v})=\mathcal{K}_n(\bar P\bar{u}|\bar Q\bar{v})\quad \forall P,Q\in S_n
\end{align}
\vspace{1cm}

$\bullet$ Let $\kappa$ be a complex parameter. We define the particle number operator $\mathcal{O}_\kappa (\bar{\mathbf{x}})$, with $\mathbf{x}$ the coordinate operator, as counting the number of particles with negative coordinate: for $\mathcal{C}(\bar{x})\in\mathcal{C}(\mathbb{R}_-^n\otimes\mathbb{R}_+^{M-n})$, $\mathcal{O}_\kappa (\bar{\mathbf{x}})\ket{\bar{x}}=\kappa^n\ket{\bar{x}}$ : 

\begin{align}
\label{PNO0}
\mathcal{O}_\kappa (\bar{\mathbf{x}})\equiv & \prod_{i=1}^{M} \mathcal{O}_\kappa (\mathbf{x}_i)
\\ \nonumber
\mathcal{O}_\kappa (\mathbf{x}_i)\equiv & \kappa^{\theta(-\mathbf{x}_i)}
\\ \nonumber
\mathbf{x}_i \ket{\bar{x}}\equiv & x_i \ket{\bar{x}}
\\ \nonumber
\theta (x)=&\left\{ 
\begin{matrix}
0 & x<0
\\
1 & x>0
\end{matrix} \right. \text{ the Heaviside function}
\end{align} 
Note that the elements of $\ket{\mathcal{C}(\mathbb{R}_-^n\times\mathbb{R}_+^{M-n})}$ for $n\in\{ 0,\cdots ,M \}$ are the maximal eigen-domains of $\mathcal{O}_\kappa (\bar{x})$, i.e. any eigen-domain of $\mathcal{O}_\kappa$ is contained in one these.
\\

\begin{note}
It is clear that the "particle number operator" as defined above is not actually returning the number of particle (on the left side of the axis) as an eigenvalue. This latter operator can however be straightforwardly define as $\mathcal{N}(\bar{\mathbf{x}})=\left. \frac{\partial \mathcal{O}_\kappa (\bar{\mathbf{x}})}{\partial \kappa} \right|_{\kappa =1}$.
\end{note}

$\bullet$ At last, for two sets $\bar x$ and $\bar u$ of cardinal $M$ and $n\leq M$ an integer, we define the shifted sets $\{ \bar x ]^n$ and $[\bar u\}^n$ by their ($i^{th}$) elements:

\begin{align}
\label{ShiftedSet}
\{ \bar{x}]^n_i\equiv \left\{ \begin{matrix}
\sum_{j=i}^n x_j \quad i\leq n
\\
\sum_{j=n+1}^i x_j \quad i>n
\end{matrix} \right.
\quad & \quad 
[\bar{u}\}^n_i\equiv \left\{ \begin{matrix}
\sum_{j=1}^i u_{j} \quad i\leq n
\\
\sum_{j=i}^{\#\bar{u}} u_{j}  \quad i>n
\end{matrix} \right.
\end{align}
We can easily check that $(\{ \bar{x}]^n,\bar{u}) = (\bar{x},[\bar{u}\}^n)$, and that  $[\bar{u}\}^n+[\bar{u}'\}^n=[\bar{u}+\bar{u}'\}^n$ (and same for $\{ \bar{x}]$). Note however that in general $\bar P [\bar{u}\}^n\neq [\bar P \bar{u}\}^n$ (and same for the action of a permutation over $\{\bar{x}]$).
\\

\textbf{The main result} of these notes can be summarized in the following theorem.

\begin{fthm}
\label{ThmPNOMESlav}
For two real sets of rapidities $\bar{u},\bar{v}\in\mathbb{R}^M$, defining the Bethe states,
\begin{align*}
\ket{\psi (\bar{u})}\equiv & \int_D \left.\psi_{\bar{u}}(\bar{x})\right|_D \ket{\bar{x}}
\\ 
= & c^{M/2}\int_{-\infty}^\infty dx_M 
\int_{-\infty}^{x_{M}} dx_{M-1} \cdots
\int_{-\infty}^{x_2} dx_1 
\sum_{P\in S_M} F(\bar P\bar{u})e^{i(\bar P\bar{u},\bar{x})}\ket{\bar{x}}
\end{align*} 
the MEPNO is expressed as 
\begin{align}
\nonumber
\bra{\psi(\bar{u})}\mathcal{O}_\kappa (\bar{\mathbf{x}})\ket{\psi(\bar{v})}= & 
\textrm{det}_{i,j}^{-1}[h^{-1}(u_i,v_j)]
\textrm{det}_{i,j}\left[  t(v_i,u_j) \frac{h(v_i,\bar u)}{h(\bar u , v_i)}\frac{h(\bar v, v_i)}{h( v_i , \bar v )}+\kappa t(u_j,v_i)   \right]
\\ \nonumber
= & \textrm{det}_{i,j}^{-1}[\frac{ic}{u_i-v_j+ic}]
\\ \nonumber
&\times\textrm{det}_{i,j}\left[  \frac{(ic)^2}{(v_i-u_j)(v_i-u_j+ic)}\ \frac{v_i-\bar u +ic}{v_i-\bar u -ic}\ \frac{ v_i -\bar v -ic}{v_i -\bar v +ic}+\kappa \frac{(ic)^2}{(v_i-u_j)(v_i-u_j-ic)}   \right]
\end{align}
\end{fthm}

This result will be proved and discussed in the last section of these notes.

\section{$\delta$-Bose gas and scalar product}
\label{SectDBGandSP}
We propose in this Section a review of the $\delta$-Bose gas problem as approached by Gaudin \cite{Gaudin}, and to reach a compact expression for the scalar product in the infinite size case. 
\\

Our system consists on $M$ indistinguishable particle on an infinite axis, with hard core interaction of intensity $c$. Its dynamics is governed by the so called non-linear Schrodinger equation:

\begin{equation}
\label{SchroDBG}
-\sum_{j=1}^M\frac{\partial ^2\psi}{\partial x_j^2}+2c\sum_{i<j}\delta (x_i-x_j)\psi=E\psi ,\quad c\in \mathbb{R},\quad \hbar=1.
\end{equation}

\subsection{Coordinate Bethe Ansatz}
For $P$ a permutation of $S_N$ and $M\in\mathbb{N}$, we define the \textit{elementary domain} $D_P=\{ \bar{x}\in\mathbb{R}^M,x_{Pi}<x_{P(i+1)} \}$. Note that $\bar{x}\in D\Leftrightarrow \bar P\bar{x}\in D_{P}$ ($D=D_{id}$). 

In the $D_P$ sector, the non-linear Schrodinger equation \eqref{SchroDBG} describes the dynamics of $M$ free particles: 

\begin{equation}
\label{SchroFree}
\left.(\Delta_N+E)\psi\right|_{D_P}=0.
\end{equation}

The full dynamical problem is specified by moreover imposing the boundary conditions: 

\begin{equation}
\label{DomainWallCond}
\left.\frac{\partial\psi}{\partial x_{i+1}}-\frac{\partial\psi}{\partial x_{i}}\right|_{x_{i+1}-x_i=0^+}=\left. c\psi\right|_{x_{i+1}=x_i}
\end{equation}
obtained from \eqref{SchroDBG} by integrating $x_i$ on an infinitesimal domain slightly surrounding $x_{i+1}$, and exploiting the complete symmetry of the wave-function under exchange of particles.
\\

The ansatz (referred to as the coordinate Bethe ansatz) consists on assuming the fundamental solution of \eqref{SchroDBG} in the domain $D$ to be a Bethe superposition of plane waves

\begin{equation}
\left.\psi_{\bar{u}}(\bar{x})\right|_D=\sum_{P\in S_N}A(\bar P\bar{u})e^{i(\bar{x},\bar P\bar{u})}
\end{equation} 
where $A$ is a set of amplitudes to be determined and $\bar{u}\in\mathbb{C}^M$ the so called set of \textit{rapidity}, obviously satisfying \eqref{SchroFree} for $E=\sum_i u_i^2$.

\begin{note}
We here assume these parameter to be complex in order to handle the case of bound states that would rise for $c>0$. These bound states are characterized by \textit{strings} of particle of rapidities linked by relations
\begin{align}
\label{BoundCondRap}
u_{i+1}-u_{i}+ic,\ c\in \mathbb{R},\ \sum_i u_i \in\mathbb{R}.
\end{align}

The existence of bound states will however be kept silent in the following, given it has no impact on our reasoning. We will consider our rapidities as being real in our manipulations, as a matter of clarity, and treat the bound state case on a more informal level. 
\end{note}
\vspace{1cm}
 
The boundary conditions \eqref{DomainWallCond} in turn reads

\begin{align}
A(\overline{P_{jj+1}P}\bar{u})=  \frac{u_{P^{-1}j}-u_{P^{-1}(j+1)} -ic}{u_{P^{-1}{j}}-u_{P^{-1}{(j+1)}} +ic}A(\bar P\bar{u}),\quad \forall P\in S_N,\ \forall j 
\end{align}
leading to the unique\footnote{up to global normalization} solution

\begin{align}
A(\bar{u})= F(\bar{u})= \prod_{i<j}\left(1+\frac{ic}{u_{i}-u_{j}}  \right)  =f^<(\bar{u},\bar{u}).
\end{align}

\begin{note}
We here obtained an expression for the wave function restricted to the fundamental domain $D$. Its expression in any other fundamental domain can straightforwardly be obtained by exploiting the complete symmetry of our wave-function under exchange of particle:
\begin{align}
\left.\psi_{\bar{u}} (\bar Q\bar{x})\right|_{D_{Q}}=\left.\psi_{\bar{u}} (\bar{x})\right|_{D}\Longrightarrow \left.\psi_{\bar{u}} (\bar{x})\right|_{D_Q}=\sum_P F(\overline{Q^{-1}P}\bar{u})e^{i(\bar{x},\bar P\bar{u})}
\end{align}

\end{note}
We can now write the Bethe states, eigenvectors of \eqref{SchroDBG},
\begin{align}
\nonumber
\ket{\psi (\bar{u})}\equiv & c^{M/2}\int_D \left.\psi_{\bar{u}}(\bar{x})\right|_D \ket{\bar{x}}
\\ \label{BE0}
= & c^{M/2}\int_{-\infty}^\infty dx_M 
\int_{-\infty}^{x_{M}} dx_{M-1} \cdots
\int_{-\infty}^{x_2} dx_1 
\sum_{P\in S_M} F(\bar P\bar{u})e^{i(\bar{x},\bar P\bar{u})}\ket{\bar{x}}
\end{align} 
where the integration running only over $D$ is consistent given the symmetry of our state, and the global normalization factor $c^{M/2}$ introduced by anticipation, as a matter of compactness.

\subsection{Scalar product and the Izergin-Korepin determinant}
We now briefly expose a method to obtain the scalar product of two Bethe states, following Gaudin (see \cite{Gaudin} Chapter $4$). 

The expression \eqref{BE0} for the Bethe functions is unfriendly in that it involves integrals of parameters over domains depending explicitly on the other integrated ones. In this regard, we shift the integrals in \eqref{BE0}, via the change of variables $\bar{x}\to \{\bar{x}]^M$ defined in \eqref{ShiftedSet}, and rewrite our Bethe states, for $\#\bar{u}=M$:

\begin{align}
\label{BEShiftedSC}
\ket{\psi (\bar{u})}=  
c^{M/2}\int_{-\infty}^\infty dx_M 
\int_{-\infty}^0 dx_{M-1} \cdots
\int_{-\infty}^0 dx_1 
\sum_{P\in S_M} F(\bar P\bar{u})e^{i(\bar{x},[\bar P\bar{u}\}^M)}\ket{\{\bar{x}]^M}
\end{align} 
where we used the identity $(\{\bar{x}]^M,\bar P\bar{u})=(\bar{x},[\bar P\bar{u}\}^M)$.
\\

Using \eqref{BEShiftedSC} and the identity $\braket{\{\bar{x}]^M|\{\bar{y}]^M}=\braket{\bar{x}|\bar{y}}=\prod_i\delta (x_i-y_i)$, we have 

\begin{align}
\label{SPInt}
\braket{\psi (\bar{u})|\psi (\bar{v})} = & 
c^M\sum_{P,Q\in S_M} F^\star(\bar P\bar{u})F(\bar Q\bar{v})
\int_{-\infty}^\infty dx_M 
\int_{-\infty}^0 dx_{M-1} \cdots
\int_{-\infty}^0 dx_1 
e^{-i(\bar{x},[\bar P\bar{u}-\bar Q\bar{v}\}^M)}
\end{align}
Integrating this expression, we obtain:

\begin{align}
\nonumber
\braket{\psi (\bar{u})|\psi (\bar{v})} = & 
c^M\sum_{P,Q\in S_M} F^\star(\bar P\bar{u})F(\bar Q\bar{v})
2\pi\delta \left( [\bar P\bar{u}-\bar Q\bar{v}\}^M_M \right)
\frac{i^{M-1}}{\prod_{i=1}^{M-1} [\bar P\bar{u}-\bar Q\bar{v}-i0\}^M_i }
\\
= & \left( \sum_i (u_i-v_i) \right)\frac{2\pi}{i}\delta \left( \sum_i (u_i-v_i) \right)
\sum_{P,Q\in S_M} F^\star(\bar P\bar{u})F(\bar Q\bar{v})
\frac{(ic)^M}{\prod_{i=1}^M [\bar P\bar{u}-\bar Q\bar{v}\}^M_i}
\end{align}
the last line being obtained noticing that $[\bar P\bar{u}-\bar Q\bar{v}\}^M_M=\sum_{i=1}^M (u_i-v_i)$, independently of $P$ and $Q$, and taking the limit $i0=0$ legally. 

\begin{note}
\label{NoteBoundConverge}
The case of bound states wouldn't here require any particular treatment (Since the rapidities are no longer real, we have to consider $\bar{u}\to\bar{u}^\star$ in the right hand side of \eqref{SPInt}). Indeed, for $P\in S_M$, one can see from \eqref{BoundCondRap} that $\mathfrak{Im} [\bar P\bar{u}^\star\}^M_i>0\Longrightarrow F(\bar P\bar{u}^\star)=0$, and same for the $Q$ part, so the corresponding integral in \eqref{SPInt} would actually vanish. The same argument will hold for the integration of \textrm{MEPNO}. 
\end{note}

One can now make use of Lemma \ref{LemmaGaudin} (See Appendix \ref{Lemmas}), due to Gaudin, and we finally obtain 
\begin{align}
\braket{\psi (\bar{u})|\psi (\bar{v})}=\frac{2\pi}{i}\left( \sum_i (u_i-v_i) \right)\delta \left( \sum_i (u_i-v_i) \right)\mathcal{K}_M(\bar{u}|\bar{v})
\end{align}

which well exhibit the orthogonality of Bethe states (note that $\mathcal{K}_M(\bar{u}|\bar{v})\propto\frac{1}{\sum_i (u_i-v_i)}$).

\begin{note}
The question of completeness of the Bethe states is not adressed here. Although very strong, the Bethe hypothesis, on which relies the ansatz, actually leads to a complete set of states. This highly non trivial fundamental feature, reflect of integrability, can be appraoched through different strategies. One can for instence cite \cite{ResIdBose} where is obtained an expression for the identity as a sum over Bethe states projectors, hence demonstrating the completeness of the Bethe states.
\end{note}  

\begin{note}\label{NoteXXXDGB}
We obtained an expression for the scalar product in the $\delta$-Bose gas in term of the so called Izergin-Korepin determinant. It is really tempting to evoke the strong similarity that this object shares with its $XXX$ cousin. 
\\
Indeed, the scalar product of two $M$-magnons states in an $XXX$ spin chain of length $2M$ can also be expressed in term of the Izergin-Korepin determinant: 
\\
For a periodic $XXX$ spin chain of inhomogeneities $\bar{\theta}$ and two sets of rapidities $\bar{v}_{I}$ and $\bar{v}_{II}$ (one of these satisfying the Bethe equations), with $\#\bar{\theta}=2\#\bar{v}_{I}=2\#\bar{v}_{II}$, the scalar product of the two corresponding Bethe states expresses as \cite{IKdetSP}
\begin{align}
\braket{\psi(\bar{v}_{I})|\psi(\bar{v}_{II})}=\mathcal{K}_{2M}(\bar{\theta}-ic/2|\{ \bar{v}_{I},\bar{v}_{II} \}).
\end{align}

While this formal similarity appearing between the scalar products in $XXX$ and in the $\delta$-Bose gas is indubitable, the role played by our parameters are in each cases very (and intriguingly) different: the formal map can be expressed as $\{ \bar{u},\bar{v} \}\to\{\bar{\theta}-ic/2,\{ \bar{v}_{I},\bar{v}_{II} \} \}$, i.e. one set of rapidities in the $\delta$-Bose gas plays the role of the inhomogeneities in $XXX$. 
\end{note}

\section{Matrix Elements of the Particle Number Operator}
\label{SectMEPNO}
We are in this section interested in the \textit{Matrix Element of the Particle Number Operator} (MEPNO).

\begin{align}
\label{PNOME0}
\mathcal{S}_\kappa^M(\bar{u}|\bar{v})\equiv\bra{\psi(\bar{u})}\mathcal{O}_\kappa (\bar{\mathbf{x}})\ket{\psi(\bar{v})},\quad \#\bar{u}=\#\bar{v}=M,
\end{align}
where $\mathcal{O}_\kappa$ is defined in \eqref{PNO0} and $\ket{\psi(\bar{v})}$ in \eqref{BE0}.
\\

\begin{note} 
Let us draft the structure of the development to follow. 
\\

Given the form of the Bethe vectors in play here, the MEPNO will involve integration over position space and two sums over permutations acting separately on the two sets of rapidity.

A first step will consist on obtaining the integrations over positions as a sum of integration over eigen-domains of $\mathcal{O}_\kappa$, i.e. summing over $n$ integrals running over $(\mathbb{R}^-)^{n}\otimes (\mathbb{R}^+)^{M-n}$. 

Then the sums over permutations and eigendomains can be translated as a sum over $2$-partitions of our sets of rapidity and sums over permutations acting on these. This is rigorously obtained in Appendix \ref{AppendixShiftingPN}. 
\\

The obtained form leads us to apply the Gaudin's Lemma \ref{LemmaGaudin}  to the two summed subsets, then providing a sum over $2$-partitions of the product of two Izergin-Korepin determinants of length $n$ and $M-n$. 
\\

After that will come a bit of rewriting in order to clean and simplify our expression, after what one could easily apply a second Lemma \ref{LemSumK}, which will leads us to a sum over partitions of a single determinant of length $M$.
\\

Theorem \ref{ThmPNOMESlav} is then obtained after a (rigorous) pinch of black magic.   
\\

\end{note}

As for the computation of the scalar product, it will be convenient to rewrite the Bethe states \eqref{BE0} as integrals over domains independent on the integrated parameters. In this case it is appropriate to write a Bethe states as a sum over eigen-domains of the particle number operator, namely domains made of product of $\mathbb{R}^\pm$. 

Such a writing is performed in Appendix \ref{AppendixShiftingPN}, so that we can rewrite the MEPNO \eqref{PNOME0} as
\begin{align}
\label{BSDecomp}
\mathcal{S}_\kappa^M(\bar{u}|\bar{v})=c^M\sum_{n=0}^M \kappa^{n} 
\prod_{i=1}^n\int_{\mathbb{R}_-} dx_i 
\prod_{i=n+1}^M \int_{\mathbb{R}_+} dx_{i}
\sum_{P,Q\in S_M} F^\star(\bar P\bar{u}) F(\bar Q\bar{v})e^{-i(\bar{x},[\bar P\bar{u}-\bar Q\bar{v}\}^n)}
\end{align}

After integration over $\bar{x}$, one obtains

\begin{align}
\label{PN}
\mathcal{S}^M_\kappa(\bar{u}|\bar{v})= & \sum_{n=0}^M \kappa^n(-)^{M-n} 
\sum_{P,Q\in S_M} \frac{(ic)^M F^\star(\bar P\bar{u}) F(\bar Q\bar{v})}{\prod_{i=1}^M[\bar P\bar{u}-\bar Q\bar{v}\}_i^n}
\end{align}

We are now to re-express the summation in  \eqref{PN}.
\\

On one hand we notice that the two terms $\prod_{i=1}^n[\bar P\bar{u}-\bar Q\bar{v}\}^n_i$ and $\prod_{i=n+1}^M[\bar P\bar{u}-\bar Q\bar{v}\}^n_i$ depend on two disjoint sets of rapidity, $\{ (\bar P \bar{u}-\bar Q \bar{v})_i \}_{i=1,\cdots , n}$ and $\{ (\bar P \bar{u}-\bar Q \bar{v})_i \}_{i=n+1,\cdots , M}$ respectively. 

On the other hand one can see that the term $F^\star (\bar P\bar{u}) F(\bar Q\bar{v})$ can be factorized in two terms depending separately on these two disjoint sets and a third crossed term depending on the normal ordered version of these sets (see Appendix \ref{ReSumPart}).  
\\

We are thus to transform the summation in \eqref{PN} as a sum over the normal ordered partitions  $\bar{u}\Rightarrow \{ \bar{u}_{I},\bar{u}_{II} \}$ and $\bar{v}\Rightarrow \{ \bar{v}_{I}, \bar{v}_{II} \}$ with $\#\bar{u}_I=\#\bar{v}_I$, and a sum over permutations acting inside each partitioned subset.  
\\

The re-summation is rigorously performed in Appendix \ref{ReSumPart}, which leads us to a re-expression of \eqref{PN} as

\begin{align}
\nonumber
\mathcal{S}^M_\kappa(\bar{u}|\bar{v}) = & \sum \kappa^{\# I} f(\bar u_{II},\bar u_{I})f(\bar v_{I},\bar v_{II})
\\ \label{PNResum2}
& \times
\sum_{P_{I},Q_{I}\in S_{\# I}} 
\frac{(ic)^{\# I}F^\star (\bar{P_I} \bar{u}_I)F (\bar{Q_I} \bar{v}_I)}{\prod_{i=1}^{\# I}[ \bar{P_{I}}\bar{u}_{I}-\bar{Q_{I}}\bar{v}_{I} \}^{\# I}_i}
\left(
\sum_{P_{II},Q_{II}\in S_{\# II}}
\frac{(ic)^{\# II}F^\star (\bar{P_{II}} \bar{u}_{II})F (\bar{Q_{II}} \bar{v}_{II})}{\prod_{i=1}^{\# II}[ \bar{P_{II}}\bar{u}_{II}-\bar{Q_{II}}\bar{v}_{II} \}^{\# II}_i}
\right)^\star
\end{align}
where the first sum runs over the normal ordered partitions  $\bar{u}\Rightarrow \{ \bar{u}_{I},\bar{u}_{II} \}$ and $\bar{v}\Rightarrow \{ \bar{v}_{I}, \bar{v}_{II} \}$ with $\#\bar{u}_I=\#\bar{v}_I$, and is made use of the notation $\# I=\#\bar{u}_{I}=\#\bar{v}_{I}$ (and same for $\# II$). 
\\

\begin{note}
This expression can be read, on an handwaving level, as a sum over the number of particle on one side of the axis, $n$, of the product of two scalar products of states living on one side of the axis or the other, weighted by $\kappa^n$. This is not so surprising given we previously were able to rewrite our states as decomposed on the operator $\mathcal{O}$'s subsectors, as in \eqref{BFShiftedPN}.     
\end{note}

We can now obviously make use of Lemma \ref{LemmaGaudin} for the two summations over permutations acting on the subsets $I$ and $II$ separately, and rewrite \eqref{PNResum2} as

\begin{align}
\label{PNResum2bis}
\mathcal{S}^M_\kappa(\bar{u}|\bar{v})
= & \sum\kappa^{\#\bar{v}_{I}}f(\bar v_{I},\bar v_{II})
\sum \mathcal{K}_{\#\bar{u}_{I}}(\bar u_{I},\bar v_{I})\mathcal{K}_{\#\bar{u}_{II}}(\bar v_{II},\bar u_{II})f(\bar u_{II},\bar u_I)
\end{align}
where we used the identity $\mathcal{K}_n^\star(\bar{u}|\bar{v})=\mathcal{K}_n(\bar{v}|\bar{u})$, and the sum over partitions split into a first sum over normally ordered partitions $\bar{v}\Rightarrow \{ \bar{v}_{I}, \bar{v}_{II} \}$ and a  second one over $\bar{u}\Rightarrow \{ \bar{u}_{I},\bar{u}_{II} \}$, with $\#\bar{u}_I=\#\bar{v}_I$.
\\ 
 
We can now straightforwardly apply Lemma \ref{LemSumK} to the second sum in \eqref{PNResum2bis} and we obtain
 
\begin{align}
\label{PNResum3}
\mathcal{S}^M_\kappa(\bar{u}|\bar{v})= & \sum_{\bar{v}\Rightarrow \{ \bar{v}_{I}, \bar{v}_{II} \}} (-\kappa)^{\# \bar v_I}f(\bar v_{I},\bar v_{II})f(\bar u,\bar v_{I}) \mathcal{K}_{M}(:\{ \bar v_{I}-c,\bar v_{II} \}:|\bar u)
\end{align}
We are here allowed to choose, as a matter of simplicity, the normal ordering for the sets in $\mathcal{K}$, given the convenient symmetry of this latter (we here simply considered $:\bar{u}:=\bar{u}$ and $:\bar{v}:=\bar{v}$). 
 \\
 
We need for the following to define the shift operator: $D^{-1}_{\bar{w}}\equiv\prod_{i=1}^{\#\bar{w}}D_{(\bar{w})_i}^{-1}$ with $D_wf(w)\equiv f(w+c)$, i.e. $D_w=e^{c\partial_w}$.
\\

\begin{note}
The normal ordering of a shifted set can be properly defined as the shifted normal ordered set.

For $\bar{v}=\{\bar v_{I},\bar v_{II}\}$ and $:\bar{v}:=\bar{v}$, $:\{ \bar v_{I}-c,\bar v_{II} \}:\ \equiv D^{-1}_{\bar{v}_I}\bar{v}$, and so $\mathcal{K}_{M}(:\{ \bar v_{I}-c,\bar v_{II} \}:|\bar u)=D^{-1}_{\bar{v}_I}\mathcal{K}_{M}( \bar v|\bar u)$ (we here sensibly consider $:\bar{v}:=\bar{v}$). 
\end{note}

After developing the Izergin-Korepin determinant in \eqref{PNResum3} according to \eqref{IKDef}, making use of the property of the different functions at play and taking care of the product over ordered indexes, we can simplify and clean a bit our expression (see Appendix \ref{AppClean}), after what we can eventually rewrite \eqref{PNResum3}
\begin{align}
\nonumber
\mathcal{S}^M_\kappa(\bar{u}|\bar{v}) = & g^<(\bar{u},\bar{u})g^>(\bar{v},\bar{v})h^{-1}(\bar{v},\bar{v})
\\ \nonumber
& \times \sum_{\bar{v}\Rightarrow \{ \bar{v}_{I}, \bar{v}_{II} \}} \kappa^{\# I}h(\bar{u},\bar{v}_{I})h(\bar{v}_I,\bar{v})h(\bar{v}_{II},\bar{u})h(\bar{v},\bar{v}_{II})D^{-1}_{\bar v_I}
\\ \label{PNResum4}
& \times \textrm{det}_{i,j}\left[t(v_i,u_j)\right]
\end{align}

Now, on one hand the element in the last line, namely the determinant, does not depends on the summed partitions in the line above. 
\\
On the other hand, the whole term in the second line can be factorized as a product of operator, using the development $\prod_{i\in\alpha} (x_i+y_i)=\sum_{\beta\subseteq\alpha  }\left[\prod_{i\in\beta}x_{i}\prod_{j\in\alpha\setminus\beta}y_{j}\right]$ (provided that $[x_{i},y_{j}]=0\ \forall i,j$). Doing so we get from \eqref{PNResum4}
 
\begin{align}
\nonumber
\mathcal{S}^M_\kappa(\bar{u}|\bar{v}) = & 
\lim\limits_{\substack{\bar{w}\to\bar{v}}} g^<(\bar{u},\bar{u})g^>(\bar{v},\bar{v})h^{-1}(\bar{v},\bar{v})
\\
& \times \prod_{i=1}^M(h(v_i,\bar u)h(\bar{w},v_i)+\kappa h(\bar{u},v_i)h(v_i, \bar w)D^{-1}_{v_i})\textrm{det}_{i,j}\left[t(v_i,u_j)\right]
\end{align}
where we simply introduce the limit $\bar{w}\to\bar{v}$ to prevent our shift operators from interacting with the other terms of the product (i.e. forcing their commutativity). 
\\

Now the last step is straightforward: we distribute the different terms of the product of operators over the different lines of the matrix, on which each act independently, and eventually obtain 
\begin{align}
\nonumber
\mathcal{S}^M_\kappa(\bar{u}|\bar{v}) = & g^<(\bar{u},\bar{u})g^>(\bar{v},\bar{v})h^{-1}(\bar{v},\bar{v})
\\ \nonumber
& \times \textrm{det}_{i,j}\left[   h(v_i,\bar u)h(\bar v, v_i)t(v_i,u_j)+\kappa h(\bar{u},v_i)h(v_i, \bar v)t(u_j,v_i)   \right]
\\ \label{MEPNOSlav}
= & \textrm{det}_{i,j}^{-1}[h^{-1}(u_i,v_j)]
\textrm{det}_{i,j}\left[\kappa t(u_j,v_i)  +t(v_i,u_j) \frac{h(v_i,\bar u)}{h(\bar u , v_i)}\frac{h(\bar v, v_i)}{h( v_i , \bar v )}   \right]
\end{align}
where we used the identity for the Cauchy determinant for the first term, and developed terms like $\prod_{i=1}^M f(x,u_i)$ over lines or column of the matrix in the last determinant.
\\

Hence Theorem \ref{ThmPNOMESlav} is proved.

\begin{note}
We are in turn here tempted to compare this expression for the MEPNO in the $\delta$-Bose gas with the expression for the scalar product in $XXX$, as obtained by N. A. Slavnov. 
\\
For $\ket{\psi (\bar{u})}_\kappa$ a Bethe state for the $\kappa$-twisted\footnote{we here refer to a diagonal twist $\begin{pmatrix}
\kappa & 0 \\ 0 & 1
\end{pmatrix}$ of the monodromy matrix in ABA.} $XXX$ spin chain and $\ket{\psi (\bar{v})}$ a Bethe vector for the non-twisted chain, the scalar product can be expressed, see \cite{Slav89SP}, in term of the so called Slavnov determinant as  

\begin{align*}
\braket{\psi(\bar{v})|\psi(\bar{u})}_\kappa=\textrm{det}_{i,j}^{-1}[h^{-1}(u_i,v_j)]
\textrm{det}_{i,j}\left[ \kappa t(u_j,v_i)- t(v_i,u_j) \frac{h(v_i,\bar u)}{h(\bar u , v_i)}\frac{h(\bar{v},v_i)}{h(v_i,\bar{v})}   \right]
\end{align*} 
which exactly corresponds to \ref{MEPNOSlav}. 
\\

While the link between the scalar product in the $\delta$-Bose gas and in $XXX$ was formally established via a one-to-one map of the spectral and inhomogeneity parameters (see Note \ref{NoteXXXDGB}), the link here seems more direct.  
\end{note}
\section{Conclusion}
The brute force approach we followed here led to the quantitative aspect of interest: a nice and compact expression for the Matrix Elements of the Particle Number Operator, which involves a determinant.       

We drew a formal link between the scalar products of Bethe states in the $\delta$-Bose gas and in $XXX$ through a mapping between rapidities of the first system and inhomogeneities of the latter. Although this link is rather obscure, it seems to try to tell us something.

On the other hand, the link between the MOPNE in the $\delta$-Bose gas and the scalar product in the $XXX$ spin chain is direct, albeit tricky. Indeed, although we may agree that a diagonal twist in $XXX$ and the Particle Number Operator are conceptually of the same nature, the lack of consistency for the concept of "number of particles on one side of a periodic spin chain" makes this connection dubious.      

These two links seem to reflect the common algebraic background shared by the $XXX$ spin chain and the $\delta$-Bose gas.   
\\

Note that the MEPNO on a segement in the periodic $XXX$ spin chain has actually already been obtained through ABA, see for instance \cite{SlavnovReview} Section 7.2, or \cite{Slav89SP}. This result however appears in a very different form than for the MEPNO in the $\delta$-Bose gas or the scalar product in the spin chain.

This important formal inadequacy simply translate the deep difference in nature between the two objects, which are the number of particle 'at the right side', and on a segment.  
\\

The $\delta$-Bose gas seems to actually be well known as a limit of its $XXX$ cousin. An approach through ABA would thus provides us with a more promising development, as it may enables us to treat the problem on the more fundamental algebraic level. In this regard, ABA often offers a powerful frame for generalization to more flavored systems. 
\\

Our result may moreover entails a practical interest, in that the MEPNO can, obviously, be used to count the number of particles lying on one side of the axis, and by extension (namely its time derivative) provides us with a concrete probe for the current at the origin of the axis. 

This may for instance be used to study the current evolution (and fluctuations) following a quench of the system, on the analytic level. 

This latter problem however brings its fistful of trouble, as the most natural approach would suggest to express our evolving states in the Bethe basis, which seems, at least at first sight, to be a very non-trivial task.

\vspace{2cm}
\subsection*{Acknowledgments}I address many thanks to Vincent Pasquier, Didina Serban and Samuel Belliard for interesting discussions and for sharing their expertise on this subject. I also want to warmly thank Veronique Terras for providing me with useful and interesting hints about the formal context of the problem.

\newpage
\appendix
\section{Appendix}

\subsection{Lemmas}
\label{Lemmas}

\begin{lemma}
\label{LemmaGaudin} 
Let $\bar{u}$ and $\bar{v}$, with $\#\bar{u}=\#\bar{v}=M$, be two sets of real parameters. Then 

\begin{align}
 \sum_{P,Q\in S_M} F^\star(\bar P\bar{u})F(\bar Q\bar{v})
\frac{(ic)^M}{\prod_{i=1}^M [P\bar{u}-Q\bar{v}\}^M_i}=\mathcal{K}_M(\bar{u}|\bar{v})
\end{align}
\end{lemma}

The proof of this Lemma can be found in \cite{Gaudin}, Appendix B.

\begin{lemma}
\label{LemSumK}
Let $\bar\gamma$, $\bar\alpha$ and $\bar{\beta}$ be sets of complex parameters with $\#\bar\alpha=m_1$, $\#\bar\beta=m_2$  and $\#\bar{\gamma}=m_1+m_2$. Then

\begin{equation}
\sum \mathcal{K}_{m_1}(\bar{\gamma}_{I}|\bar{\alpha})\mathcal{K}_{m_2}(\bar{\beta}|\bar{\gamma}_{II})f(\bar{\gamma}_{II},\bar{\gamma}_{I})
=(-)^{m_1}f(\bar{\gamma},\bar{\alpha})\mathcal{K}_{m_1+m_2}(\{\bar{\alpha}-c,\bar{\beta} \}|\bar{\gamma})
\end{equation}

where summation is made over the \textit{normal ordered} partitions $\bar{\gamma}\Rightarrow\{ \bar{\gamma}_{I},\bar{\gamma}_{II} \}$, with $\#\bar{\gamma}_{I}=\#\bar{\alpha}$.
\end{lemma}

The proof for this lemma can be found in \cite{BelliardABAforSPsu3}.

\subsection{Projecting the integrals over \textit{Particle Number Operator} maximal eigen-domains.}
\label{AppendixShiftingPN}

Let's define the domain $D^n\subset D$ as the domain of (strictly) ordered positions with $n$ of these being negatives:

\begin{align}
D^n=\{ \bar{x}\in\mathbb{R}^M|x_1<\cdots <x_n<0<x_{n+1}<\cdots <x_M \}
\end{align}
These domains are actually the ordered \textit{particle number operator}'s eigen-domains: $\mathcal{O}_\kappa (\bar{\mathbf{x}}) \ket{D^n}=\kappa^n\ket{D^n}$. 

Then the Bethe functions \eqref{BE0} can be rewritten as a sum over integrated domain $D^n$   
\begin{align}
\nonumber
\ket{\psi (\bar{u})}= & c^{M/2}\sum_{n=0}^M\int_{D^n}d\bar{x} \sum_{P\in S_M} F(\bar P\bar{u})e^{-i(\bar P\bar{u},\bar{x})}\ket{\bar{x}}
\\ 
\nonumber
= & c^{M/2}\sum_{n=0}^M
\int_{-\infty}^0 dx_n 
\int_{-\infty}^{x_n} dx_{n-1}
\cdots \int_{-\infty}^{x_2} dx_{1}
\int_{0}^{\infty} dx_{n+1}
\int_{x_{n+1}}^{\infty} dx_{n+2}
\cdots
\int_{x_{M-1}}^{\infty} dx_{M}
\\ \label{BFShiftedPN0}
& \times\sum_{P\in S_M} F(\bar P\bar{u})e^{-i(\bar P\bar{u},\bar{x})}\ket{\bar{x}}
\end{align}

It will appear important, for the computations of the desired quantity, to shift the integration domains to domains independent of the integrated positions. Thus, shifting the integration boundaries in \eqref{BFShiftedPN0} to $\mathbb{R}^\pm$, one can write the Bethe function

\begin{align}
\nonumber
\ket{\psi (\bar{u})}= & c^{M/2}\sum_{n=0}^M  
\prod_{i=1}^n\int_{\mathbb{R}^-} dx_i 
\prod_{i=n+1}^M \int_{\mathbb{R}^+} dx_{i}
\sum_{P\in S_M} F(\bar P\bar{u})e^{-i(\{\bar{x}]^n,\bar P\bar{u})}\ket{\{\bar{x}]^n}
\\ \label{BFShiftedPN}
= & c^{M/2}\sum_{n=0}^M 
\prod_{i=1}^n\int_{\mathbb{R}^-} dx_i 
\prod_{i=n+1}^M \int_{\mathbb{R}^+} dx_{i}
\sum_{P\in S_M} F(\bar P\bar{u})e^{-i(\bar{x},[\bar P\bar{u}\}^n)}\ket{\{\bar{x}]^n}
\end{align}
We here used the identity $(\{\bar{x}]^n,\bar P\bar{u})=(\bar{x},[\bar P\bar{u}\}^n)$.

\subsection{Resummation over $2$-partitions}
\label{ReSumPart}
For any $P$ and $Q$ elements of $S_M$ and $n\leq M$ an integer, we uniquely define 
\begin{align}
\nonumber
\bar P\bar{u} = & \{\bar{P_{I}}\bar{u}_{I},\bar{P_{II}}\bar{u}_{II}\}
\\ \nonumber
\bar{u}_{I}= & :\bar{P_{I}}\bar{u}_{I}:
\\ \nonumber
\bar{u}_{II}= & :\bar{P_{II}}\bar{u}_{II}:
\\ \nonumber
&
\\
\#\bar{u}_{I}=&n=\#\bar{v}_{I}
\\ \nonumber
&
\\ \nonumber
\bar Q\bar{v} = & \{\bar{ Q_{I}}\bar{v}_{I},\bar{Q_{II}}\bar{v}_{II}\}
\\ \nonumber
\bar{v}_{I}= & :\bar{Q_{I}}\bar{u}_{I}:
\\ \nonumber
\bar{v}_{II}= & :\bar{Q_{II}}\bar{v}_{II}:
\end{align}

Now, one can easily show that for $\bar{w}=\{ \bar{w}_{I},\bar{w}_{II} \}$, with $\#\bar{w}=M$ and $\#\bar{w}_{I}=n$, we have

\begin{align}
\prod_{i=1}^{n}[ \bar{w} \}^{n}_i
= & \prod_{i=1}^{n}[ \bar{w_{I}} \}^{n}_i
\\
\prod_{i=n+1}^{M} [\bar{w} \}^n_i
= & \prod_{i=1}^{M-n} [\bar{w}_{II} \}^0_i
\end{align}

and that

\begin{align}
F(\bar{w}) & =F(\bar{w}_{I})F(\bar{w}_{II})f(\bar{w}_{I},\bar{w}_{II})
\\
f(\bar{w}_{I},\bar{w}_{II}) &= f(\bar P\bar{w}_{I},\bar Q\bar{w}_{II})\quad\forall\ P,Q.
\end{align}

Combining the preceding matchings and properties, one can effectively rewrite \eqref{PN} as

\begin{align}
\nonumber
\mathcal{S}^M_\kappa(\bar{u},\bar{v}) = & \sum\sum \kappa^{\#\bar u_I}(-)^{\#\bar u_{II}} f^\star(\bar u_{I},\bar u_{II})f(\bar v_{I},\bar v_{II})
\\ \label{PNResum}
& \times
\sum_{P_{I},Q_{I}} 
\frac{(ic)^{\#\bar u_{I}}F^\star (\bar{P_I} \bar{u}_I)F (\bar{Q_I} \bar{v}_I)}{\prod_{i=1}^{\#\bar{u}_{I}}[ \bar{P_{I}}\bar{u}_{I}-\bar{Q_{I}}\bar{v}_{I} \}^{\#\bar{u}_{I}}_i}
\times\sum_{P_{II},Q_{II}}
\frac{(ic)^{\#\bar u_{II}}F^\star (\bar{P_{II}} \bar{u}_{II})F (\bar{Q_{II}} \bar{v}_{II})}{\prod_{i=1}^{\#\bar{u}_{II}}[ \bar{P_{II}}\bar{u}_{II}-\bar{Q_{II}}\bar{v}_{II} \}^0_i}
\end{align}
where the first pair of sum run over the normal ordered partitions  $\bar{u}\Rightarrow \{ \bar{u}_{I},\bar{u}_{II} \}$ and $\bar{v}\Rightarrow \{ \bar{v}_{I}, \bar{v}_{II} \}$, with $\#\bar{u}_I=\#\bar{v}_I$, and $P_I, P_{II},Q_I$ and $Q_{II}$ acts on $\bar{u}_{I},\bar{u}_{II},\bar{v}_{I}$ and $\bar{v}_{II}$ respectively. 
\\

Now one last bit of rewriting: using

\begin{align}
f^\star(u,v)=&f(v,u)
\\
\Rightarrow F^\star (\bar{w})=&F(\bar T\bar{w})
\\
\prod_{i=1}^{\#\bar{w}}[\bar{w} \}^0_i
=& \prod_{i=1}^{\#\bar{w}}[\bar T\bar{w}\}^{\#\bar{w}}_i
\end{align} 
where $T$ is defined as the "mirror" permutation: $Ti=n-i+1$, with $n=\#\bar{w}$,

one can rewrite \eqref{PNResum} as
\begin{align}
\nonumber
\mathcal{S}^M_\kappa(\bar{u},\bar{v}) = & \sum\sum \kappa^{\#\bar u_I} f(\bar u_{II},\bar u_{I})f(\bar v_{I},\bar v_{II})
\\ \nonumber
& \times
\sum_{P_{I},Q_{I}} 
\frac{(ic)^{\#\bar u_{I}}F^\star (\bar{P_I} \bar{u}_I)F (\bar{Q_I} \bar{v}_I)}{\prod_{i=1}^{\#\bar{u}_{I}}[ \bar{P_{I}}\bar{u}_{I}-\bar{Q_{I}}\bar{v}_{I} \}^{\#\bar{u}_{I}}_i}
\left(
\sum_{P_{II},Q_{II}}
\frac{(ic)^{\#\bar u_{II}}F^\star (\bar T\bar{P_{II}} \bar{u}_{II})F (\bar T \bar{Q_{II}} \bar{v}_{II})}{\prod_{i=1}^{\#\bar{u}_{II}}[ \bar{T}(\bar{P_{II}}\bar{u}_{II}-\bar{Q_{II}}\bar{v}_{II}) \}^{\#\bar{u}_{II}}_i}
\right)^\star
\\ \nonumber
= & \sum\sum \kappa^{\#\bar u_I} f(\bar u_{II},\bar u_{I})f(\bar v_{I},\bar v_{II})
\\
& \times
\sum_{P_{I},Q_{I}} 
\frac{(ic)^{\#\bar u_{I}}F^\star (\bar{P_I} \bar{u}_I)F (\bar{Q_I} \bar{v}_I)}{\prod_{i=1}^{\#\bar{u}_{I}}[ \bar{P_{I}}\bar{u}_{I}-\bar{Q_{I}}\bar{v}_{I} \}^{\#\bar{u}_{I}}_i}
\left(
\sum_{P_{II},Q_{II}}
\frac{(ic)^{\#\bar u_{II}}F^\star (\bar{P_{II}} \bar{u}_{II})F ( \bar{Q_{II}} \bar{v}_{II})}{\prod_{i=1}^{\#\bar{u}_{II}}[ \bar{P_{II}}\bar{u}_{II}-\bar{Q_{II}}\bar{v}_{II} \}^{\#\bar{u}_{II}}_i}
\right)^\star
\end{align}

\subsection{Simplifying and cleaning}
\label{AppClean}

Using the definition \eqref{IKDef} of $\mathcal{K}_n(\bar{x}|\bar{y})$ we can develop the Izergin-Korepin determinent 
\begin{align}
\nonumber
f(\bar v_{I},\bar v_{II})f(\bar u,\bar v_{I}) D^{-1}_{\bar{v}_I} \mathcal{K}_{M}(\bar{v}|\bar u)
= & f(\bar v_{I},\bar v_{II})f(\bar u,\bar v_{I})
\\ \nonumber
& \times
g^<(\bar v_{I},\bar v_{I})g^<(\bar v_{II},\bar v_{II})g^<(\bar v_{I}-c,\bar v_{II})g^<(\bar v_{II},\bar v_{I}-c)g^>(\bar u , \bar u)
\\ \label{AppClean1}
& \times h(\bar v_{I}-c,\bar u)h(\bar v_{II},\bar u)
D^{-1}_{\bar v_I}det_{i,j}\left[t(v_i,u_j)\right]
\end{align}

Now using the properties (from \eqref{RatFct1}-\eqref{RatFct2})

\begin{align}
f(x,y)= & h(x,y)g(x,y)
\\
g(x,y)=& -g(y,x)
\\
g(x,y-c)= & h^{-1}(x,y)\Rightarrow g^<(\bar{v}_{II},\bar{v}_{I}-c) =\left(h^<(\bar{v}_{II},\bar{v}_{I})\right)^{-1}
\\
g(x-c,y)= & -h^{-1}(y,x)\Rightarrow g^<(\bar{v}_{I}-c,\bar{v}_{II}) =\left(h^>(\bar{v}_{II},\bar{v}_{I})\right)^{-1}(-)^{\xi (\bar{v}_I,\bar{v}_{II}) }
\\
h(x-c,y)= & g^{-1}(x,y)
\end{align}

where we defined $(-)^{\xi (\bar{v}_I,\bar{v}_{II})}\equiv\prod\limits_{\substack{v_i\in\bar{v}_{I} \\ v_j\in\bar{v}_{II} \\ i<j }}(-1)$, we can write \eqref{AppClean1} as

\begin{align}
\nonumber
f(\bar v_{I},\bar v_{II})f(\bar u,\bar v_{I}) D^{-1}_{\bar{v}_I}\bar{v} \mathcal{K}_{M}(\bar{v}|\bar u) = & 
h(\bar v_{I},\bar v_{II})g(\bar v_{I},\bar v_{II})h(\bar u,\bar v_{I})g(\bar u,\bar v_{I})
\\ \nonumber
& \times  g^<(\bar v_{I},\bar v_{I})g^<(\bar v_{II},\bar v_{II}) \left(h^>(\bar{v}_{II},\bar{v}_{I})\right)^{-1}(-)^{\xi (\bar{v}_I,\bar{v}_{II}) }\left(h^<(\bar{v}_{II},\bar{v}_{I})\right)^{-1}
\\ \nonumber
& \times g^>(\bar u , \bar u)
\\ \nonumber
& \times  
(-)^{\#\bar{v}_I M}g^{-1}(\bar u,\bar v_{I})h(\bar v_{II},\bar u)
\\
& \times  D^{-1}_{\bar v_I}det_{i,j}\left[t(v_i,u_j)\right].
\end{align}

Now notice that $g(\bar v_{I},\bar v_{II})g^<(\bar v_{I},\bar v_{I})g^<(\bar v_{II},\bar v_{II})(-)^{\xi (\bar{v}_I,\bar{v}_{II}) }=g^<(\bar{v},\bar{v})(-)^{\# I\# II}$ ($\xi (\bar{v}_{II},\bar{v}_{I})+\xi (\bar{v}_I,\bar{v}_{II}) =\# I\# II$) and that $\frac{h(\bar{v}_{I},\bar{v}_{II})}{h(\bar{v}_{II},\bar{v}_{I})}=\frac{h(\bar{v}_{I},\bar{v})}{h(\bar{v},\bar{v}_{I})}=\frac{h(\bar{v},\bar{v}_{II})h(\bar{v}_{I},\bar{v})}{h(\bar{v},\bar{v})}$, so \eqref{AppClean1} eventually rewrites
\begin{align}
\nonumber
f(\bar v_{I},\bar v_{II})f(\bar u,\bar v_{I}) D^{-1}_{\bar{v}_I}\bar{v} \mathcal{K}_{M}(\bar{v}|\bar u)  
=
& g^<(\bar{u},\bar{u})g^>(\bar{v},\bar{v})h^{-1}(\bar{v},\bar{v})
\\ \nonumber
& \times h(\bar{u},\bar{v}_{I})h(\bar{v}_I,\bar{v})h(\bar{v}_{II},\bar{u})h(\bar{v},\bar{v}_{II})(-)^{\# I \# II}(-)^{M\# I}
\\
& \times D^{-1}_{\bar v_I}det\left[t(v_i,u_j)\right]
\end{align}

Just notice that $(-)^{\# I(M-\# II)}=(-)^{\# I^2}=(-)^{\# I}$.

\newpage

\bibliography{ms}

\end{document}